%% file: 00_main.tex
\documentclass[12pt, authoryear, letterpaper]{article}

\input{styles/structure}

\title{\Large \textbf{What Quality Engineers Need to Know About Generative AI: Structured Data Extraction}}

\author[1]{Fadel M. Megahed}
\author[2]{Ying-Ju Chen}
\author[3]{Yamin Dahwich}
\author[4]{Zhao-Wei Chen}
\author[5]{Lucas K. Hung}
\author[1]{Arthur Carvalho}
\author[1]{L. Allison Jones-Farmer}
\author[6]{Ibrahim Yousif}
\author[7,*]{Inez M. Zwetsloot}

\affil[1]{Farmer School of Business, Miami University, Oxford, OH 45056, USA}
\affil[2]{College of Arts and Sciences, University of Dayton, Dayton, OH 45469, USA}
\affil[3]{Trinity College, University of Cambridge,  Cambridge CB2 1TQ, United Kingdom}
\affil[4]{School of Business Administration, University of Dayton, OH 45469, USA}
\affil[5]{School of Engineering, University of Dayton, Dayton, OH 45469, USA}
\affil[6]{Department of Applied Computing and Engineering Technology, Miami University, Hamilton, OH 45011, USA}
\affil[7]{Amsterdam Business School, University of Amsterdam, Amsterdam, The Netherlands}
\affil[*]{Corresponding author. Email: \href{mailto:i.m.zwetsloot@uva.nl}{i.m.zwetsloot@uva.nl}}

\date{}

\begin{document}

\maketitle

% \bigskip

\vspace{-1.5\baselineskip}

\begin{abstract}
    \noindent In this expository paper, we explain how quality engineers can use large language models (LLMs) and vision-language models (VLMs) for structured data extraction. We introduce the relevant concepts, discuss structured extraction from a measurement-system perspective, and demonstrate its implementation in R and Python. We then present two freely available web-based applications (apps): (a) extracting information from automotive recall notices and (b) monitoring quality engineering research on arXiv. To ensure the accuracy of our AI-driven structured extraction, human evaluators validated the results. In the first app, four reviewers assessed information extracted from 30 recall notices using two models, with 98.6\% of the 720 field-level ratings judged fully correct. In the second app, 21 authors evaluated information extracted from 45 of their arXiv papers. All paper summaries were rated mostly correct or correct, although the evaluations identified problems with some classification fields and mathematical expressions. These results demonstrate the practical potential of structured data extraction while emphasizing the need to validate the extracted information. We make both apps and their source code publicly available so that readers can adapt the approach to their own applications.

\end{abstract}

\noindent \keywords{Data quality; generative AI; JSON schema; large language models; structured data extraction; unstructured data; vision-language models}

\section{Introduction}
\label{sec:intro}
\input{01_intro}

\section{Background}
\label{sec:back}
\input{02_background}

\section{Structured Data Extraction with LLMs/VLMs}
\label{sec:structured}
\input{03_structured_extraction}

\section{Application 1: Recall Notice Data Extraction}
\label{sec:app1}
\input{04_application_1}

\section{Application 2: Monitoring QE Research on arXiv}
\label{sec:app2}
\input{05_application_2}

\section{Discussion}
\label{sec:disc}
\input{06_discussion}

\section{Concluding Remarks}
\label{sec:conc}
\input{07_concluding_remarks}

\section*{Reproducibility Materials}
The code and supporting materials for this paper are available at \url{https://github.com/fmegahed/qe_genai_extraction}. The repository includes the NHTSA recall sample data, R scripts and results for model comparisons, and materials for the human evaluations, including the review application and a short demonstration video. The deployed versions of both applications are available through the links provided in their respective sections. The evaluated versions of Applications 1 and 2 are tagged in their respective repositories on Hugging Face and GitHub. Furthermore, the factsheets that the authors rated are archived unchanged in the second repository.

\section*{Use of Generative AI Tools}
We used Claude Code to develop the two apps and their associated evaluation tools through AI-assisted coding (often known as ``vibe coding"), as described in our GitHub repository. During manuscript preparation, we used ChatGPT, Claude, Grammarly, and Writefull to improve grammar, edit and shorten text, and reduce repetition across sections. We also used Claude Code and ChatGPT to improve the formatting and presentation of the \LaTeX \ figures.

\section*{Acknowledgments}
The authors thank the 21 authors of arXiv papers who accepted our factsheet review invitations, provided feedback, and completed our survey. We also thank Maria Weese and {\"O}zge S{\"u}rer for their feedback on earlier versions of \texttt{QE ArXiv}. We acknowledge Miami University's \textit{Raymond E. Glos Professorship in Business}, \textit{Dinesh \& Ila Paliwal Innovation Chair}, and \textit{Van Andel Endowed Professorship in Business Analytics} for supporting the AI usage costs.

\printbibliography

\end{document}

%% file: styles/structure.tex
\usepackage[utf8]{inputenc}

\usepackage[export]{adjustbox}
\usepackage{algorithm}
\usepackage[noend]{algpseudocode}
\usepackage{amsmath}
\usepackage{amssymb}
\usepackage{animate}
\usepackage{appendix}
\usepackage{array}
\usepackage{authblk}
\usepackage{bm}
\usepackage{booktabs}
\usepackage{caption}
\usepackage{color}
\usepackage{colortbl}
\usepackage{changepage}
\usepackage{chemformula}
\usepackage{comment}
\usepackage{csquotes}
\usepackage{csvsimple}
\usepackage{etoolbox}
\usepackage{enumitem}
\usepackage{float}
\usepackage{framed}
\usepackage{fontawesome5}
\usepackage[margin=1in]{geometry}
\usepackage{graphicx}
\definecolor{links}{HTML}{0074CC}
\usepackage[colorlinks=true, allcolors=links]{hyperref}
\usepackage{lineno}
\usepackage{longtable}
\usepackage{lscape}
\usepackage{listings}
\usepackage{makecell}
\usepackage{marginnote}
\usepackage{mathrsfs}
\usepackage{mathtools}
\usepackage{mdframed}
\usepackage{multicol}
\usepackage{multirow}
\usepackage{pdflscape}
\usepackage{rotating}
\usepackage{tabularx}
\usepackage{tabulary}
\usepackage{tikz}
\usepackage[textsize=scriptsize]{todonotes}
\usepackage{setspace}
\usepackage{soul}
\usepackage{subcaption}
\usetikzlibrary{shapes, arrows, positioning, fit}
\usepackage{tabularray}
\usepackage{wrapfig}
\usepackage{xargs}
\usepackage[normalem]{ulem}

\usepackage{epstopdf}
\AppendGraphicsExtensions{.tif}
\PassOptionsToPackage{hyphens}{url}\usepackage{hyperref}

\usepackage[authordate, natbib=true, backend=biber]{biblatex-chicago}
\definecolor{sandstorm}{rgb}{0.93, 0.84, 0.25}
\definecolor{miamired}{HTML}{C3142D}
\definecolor{light-tan}{HTML}{EDECE2}
\definecolor{mabordeaux}{HTML}{C41230}
\definecolor{magold}{HTML}{EFDB72}
\definecolor{maslate}{HTML}{3E5468}
\definecolor{magray}{HTML}{666666}

\definecolor{cb1}{RGB}{166, 206, 227}  % Light blue
\definecolor{cb2}{RGB}{31, 120, 180}   % Dark blue
\definecolor{cb3}{RGB}{178, 223, 138}  % Light green
\definecolor{cb4}{RGB}{51, 160, 44}    % Dark green
\definecolor{cb5}{RGB}{251, 154, 153}  % Light red
\definecolor{cb6}{RGB}{227, 26, 28}    % Dark red
\definecolor{cb7}{RGB}{253, 191, 111}  % Light orange
\definecolor{cb8}{RGB}{255, 127, 0}    % Dark orange
\definecolor{cb9}{RGB}{202, 178, 214}  % Light purple
\definecolor{cb10}{RGB}{106, 61, 154}  % Dark purple
\definecolor{cb11}{RGB}{255, 255, 153} % Light yellow
\definecolor{cb12}{RGB}{177, 89, 40}   % Brown 
\definecolor{cb13}{RGB}{220, 220, 220} % Light gray
\definecolor{cb14}{RGB}{80, 80, 80}    % Dark gray

\usetikzlibrary{positioning,shapes,arrows,arrows.meta,shadows,fit,backgrounds,calc}

\providecommand{\keywords}[1]{\textbf{{Keywords:}} #1}

\newcommand{\instructions}[1]{}

\newcolumntype{C}[1]{>{\centering\arraybackslash}p{#1}}

\providecommand{\keywords}[1]
{
  \noindent \small	
  \textbf{Keywords: } #1
}

\makeatletter
\newcommand*{\centerfloat}{%
  \parindent \z@
  \leftskip \z@ \@plus 1fil \@minus \textwidth
  \rightskip\leftskip
  \parfillskip \z@skip}
\makeatother

\AtBeginEnvironment{tabular}{\footnotesize} % Makes all tables at footnote size
\definecolor{miamired}{HTML}{C3142D}

\definecolor{cb1}{RGB}{166, 206, 227}  % Light blue
\definecolor{cb2}{RGB}{31, 120, 180}   % Dark blue
\definecolor{cb3}{RGB}{178, 223, 138}  % Light green
\definecolor{cb4}{RGB}{51, 160, 44}    % Dark green
\definecolor{cb5}{RGB}{251, 154, 153}  % Light red
\definecolor{cb6}{RGB}{227, 26, 28}    % Dark red
\definecolor{cb7}{RGB}{253, 191, 111}  % Light orange
\definecolor{cb8}{RGB}{255, 127, 0}    % Dark orange
\definecolor{cb9}{RGB}{202, 178, 214}  % Light purple
\definecolor{cb10}{RGB}{106, 61, 154}  % Dark purple
\definecolor{cb11}{RGB}{255, 255, 153} % Light yellow
\definecolor{cb12}{RGB}{177, 89, 40}   % Brown 
\definecolor{cb13}{RGB}{220, 220, 220} % Light gray
\definecolor{cb14}{RGB}{80, 80, 80}    % Dark gray

\usetikzlibrary{positioning,shapes,arrows,shadows,fit,backgrounds}

%% file: 01_intro.tex
Warranty claims and customer complaints often describe product failures in free-form text rather than as variables that can be directly analyzed. This is one of the practical challenges associated with increasingly large and complex data sources in quality engineering \citep{hong2018big,megahed2015statistical,weese2016statistical}. For example, a warranty claim may identify the affected component, describe the failure, and document the service action. To identify the most common failure modes, we would need to extract and classify this information from each claim, combine the results into a structured dataset, and summarize them using a Pareto chart. Performing this extraction manually across thousands of claims would be time-consuming. Furthermore, consistent coding would require clear guidelines, particularly when the same failure is described differently across claims.

Previous studies have used text mining and natural language processing (NLP) to extract and classify information from consumer complaints and automotive defect records \citep{ghazizadeh2014text,francis2023smartxt,uglanov2024nlp}. While these approaches can reduce manual coding, they require developing extraction rules and/or training NLP models for a specific task. Large language models (LLMs) and vision-language models (VLMs) provide an alternative for performing \textit{structured data extraction}. Specifically, one can specify the fields of interest and describe what they want to extract in plain English (or another language of their choice). Using this information, the LLM/VLM processes the inputted records and extracts the defined fields in a structured format \citep{openai2025structured}. 

In this expository paper, we demonstrate how quality engineers can use LLMs and VLMs for structured data extraction. We first introduce the relevant concepts and discuss when this approach may be useful. We then provide implementation examples in R and Python, followed by two applications: (a) extracting information from automotive recall notices and (b) monitoring new quality engineering research on arXiv. Both applications are freely available online, and we report the results of human evaluations of their extracted outputs.

%% file: 02_background.tex
\subsection{Structured Data Extraction as a Measurement System}

The process of using AI for structured data extraction hinges on an LLM/VLM and an output schema, which together can be conceptualized as a measurement system for extracting variables. In this analogy, the measurement system can take the following inputs:

\begin{enumerate}[nosep, label=(\arabic*)]
    \item input information that contains the data that we want to extract (e.g., a complaint narrative, a scanned form, a PDF, or a photograph);
    \item a short prompt describing the extraction task;
    \item a \emph{schema} that names the fields to be extracted, sets each field's data type, and then defines each field in plain language; and
    \item optionally, inputting a few worked examples to provide the model with better context.
\end{enumerate}

\noindent The AI model then produces a single structured record with the extracted variables in their pre-specified data types. For a single record, the resulting output is now a structured dataset that can be appended to a data frame row. This pipeline is described in detail in Section~\ref{sec:structured}.

The extraction system can be evaluated on three dimensions: (a) correctness, which assesses if the extracted data accurately reflect the source document; (b) repeatability, which measures if repeated extractions from the same document lead to consistent results; and (c) reproducibility, which determines if results are consistent when different LLMs/VLMs are used. In our two applications, we focus on correctness by asking human reviewers to evaluate the extracted information against the source documents. Although we compare two models in Application 1, we do not conduct a formal study of repeatability or reproducibility. The interested reader is referred to \citet{megahed2025reliable} for detailed coverage of these issues in LLM-based text classification applications.

\subsection{A Plain-Language Account of LLMs and VLMs}

Large language models are generally based on transformer architectures \citep{vaswani2017attention}. They are pretrained on large collections of text \citep{brown2020language}, and many are subsequently refined to follow user instructions \citep{ouyang2022training}. For our purposes, an important capability is \emph{in-context learning}, where the model performs a task based on instructions and, optionally, a few examples provided by the user, without requiring additional training \citep{brown2020language}. This allows us to define extraction tasks in plain English rather than developing a separate model for each application.

Quality engineering records often contain information in visual forms, including photographs, inspection images, and scanned documents. Vision-language models (VLMs) extend the capabilities of LLMs by allowing images to be used as inputs. For example, a VLM can extract information from scanned inspection forms or image-based PDFs, as well as interpret visual information in photographs. Beyond text extraction, related vision-language approaches have also been investigated for manufacturing image inspection using visual characteristics of products rather than text within images \citep{megahed2026adapting}. For a more detailed discussion of LLMs and their applications in statistics, we refer readers to \citet{ji2026overview}.

\subsection{Terms Quality Engineers Need for Structured Data Extraction}

We presume that readers may have varied AI-related backgrounds. Hence, Table~\ref{tab:terms} defines the key AI terms used throughout our paper.

\begin{table}[htb!]
\centering
\caption{Key generative AI terms for quality engineers.}
\vspace{-0.5\baselineskip}
\label{tab:terms}
\setlength{\tabcolsep}{5pt}
\begin{tabular}{p{2.3cm}p{6.47cm}p{7.1cm}}
\toprule
\textbf{Term} & \textbf{Plain-English meaning} & \textbf{Why it matters for data extraction} \\
\midrule
LLM & A model that processes and generates text in response to instructions & Can extract information from complaint narratives or reports \\ \addlinespace[3pt]
VLM & A model that processes both images and text & Can extract information from scanned forms, photographs, or image-based PDFs \\ \addlinespace[3pt]
Prompt & Instructions provided to the model & Specifies what information should be extracted \\ \addlinespace[3pt]
Schema & Field names, data types, and descriptions defining the required output & Similar to a coding sheet with operational definitions for each variable \\ \addlinespace[3pt]
Token/context window & Tokens are units of text; the context window limits how many tokens the model can process in one request & Affects how much information can be processed at once and the associated cost \\ \addlinespace[3pt]
Structured JSON output & A machine-readable record containing the fields specified in the schema & Can be converted into a row in a data frame \\ \addlinespace[3pt]
Hallucination & An incorrect or unsupported output generated by the model & Extracted values may require verification against the source document \\
\bottomrule
\end{tabular}
\end{table}

One limitation of LLMs and VLMs is that they may generate information that is not supported by the source document, commonly referred to as \emph{hallucination} \citep{huang2025survey}. 
For structured data extraction, we are particularly concerned with whether the extracted values are supported by the source document (\emph{faithfulness}) \citep{huang2025survey}. This is closely related to \emph{epistemic accuracy}, which emphasizes that producing correct results depends on having access to the necessary evidence and using it appropriately \citep{hamilton2026token}. For example, if a maintenance report states that a motor overheated but does not identify the cause, the model should report the root cause as unknown rather than assume that a cooling system failure was responsible.

\subsection{Relevance to Quality Engineering}

Interest in structured data extraction is part of a broader shift across engineering and statistics. Recent surveys document the rapid adoption of GenAI in design, manufacturing, and quality control \citep{nguyen2025review}; reviews in applied statistics describe an increasingly complementary relationship between statistical methods and AI tools \citep{min2026applied}; and an industry--academic--government panel on the future of statistical engineering argues that the discipline should embrace data science and AI as core capabilities \citep{anderson2022statistical}. Within quality engineering specifically, recent work has begun integrating LLMs and VLMs with quality control \citep{megahed2024generative, megahed2024introducing, megahed2026adapting}. In this paper, we introduce a different use case for these models: extracting structured information from documents/images to enable subsequent analyses.

This approach is most useful when records are numerous and text-heavy, the fields of interest can be defined in advance, and a sample of the extracted records can be checked against the source documents. It is less useful when the task is small enough to code reliably by hand, simple deterministic rules would suffice, the required information is not present in the source documents, or errors could have serious consequences and no review process is planned. When documents contain proprietary or sensitive information, organizations should also consider whether they can be sent to an external model provider (See Section~\ref{subsec:improvements}).

%% file: 03_structured_extraction.tex
\begin{figure}[htb!]
    \centering
    \begin{tikzpicture}[
        node distance=1.1cm and 1.4cm,
        >={Stealth[length=3mm]},
        every node/.style={font=\small},
        inputnode/.style={
            rectangle, rounded corners=4pt, draw=maslate, line width=0.8pt,
            fill=white, minimum height=1.0cm, minimum width=3cm,
            text centered, font=\footnotesize
        },
        processnode/.style={
            rectangle, rounded corners=4pt, draw=mabordeaux, line width=0.8pt,
            fill=white, minimum height=1.0cm, minimum width=3cm,
            text centered, font=\footnotesize
        },
        outputnode/.style={
            rectangle, rounded corners=4pt, draw=black, line width=0.8pt,
            fill=white, minimum height=1.0cm, minimum width=3cm,
            text centered, font=\footnotesize
        },
        schemanode/.style={
            rectangle, rounded corners=4pt, draw=magold!80!black, line width=0.8pt,
            fill=white, minimum height=1.0cm, minimum width=3cm,
            text centered, font=\footnotesize
        },
        arr/.style={->, line width=0.7pt, color=magray}
    ]
        % --- Input column ---
        \node[inputnode] (text) {Free-form Text};
        \node[inputnode, below=0.45cm of text] (pdf) {PDF Document};
        \node[inputnode, below=0.45cm of pdf] (img) {Image / Scan};

        % Brace
        \node[fit=(text)(img), inner sep=0pt] (inputs) {};

        % --- Schema ---
        \node[schemanode, right=2.2cm of pdf, yshift=1.4cm] (schema) {
            \begin{tabular}{@{}c@{}}
                User-Defined\\[-2pt] Schema
            \end{tabular}
        };

        % --- LLM ---
        \node[processnode, right=2.2cm of pdf] (llm) {LLM / VLM};

        % --- Structured output ---
        \node[outputnode, right=2.2cm of llm, yshift=0.7cm] (json) {JSON Record};
        \node[outputnode, right=2.2cm of llm, yshift=-0.7cm] (df) {Data Frame};

        % --- Arrows from inputs to LLM ---
        \draw[arr] (text.east) -- ++(0.6,0) |- (llm.west);
        \draw[arr] (pdf.east) -- (llm.west);
        \draw[arr] (img.east) -- ++(0.6,0) |- (llm.west);

        % --- Schema to LLM ---
        \draw[arr, color=magold!80!black] (schema.south) -- (llm.north);

        % --- LLM to JSON, JSON to Data Frame ---
        \draw[arr] (llm.east) -- ++(0.6,0) |- (json.west);
        \draw[arr] (json.south) -- (df.north);

        % --- Labels ---
        \node[above=0.15cm of text, font=\footnotesize\bfseries, color=maslate] {Unstructured Inputs};
        \node[above=0.15cm of schema, font=\footnotesize\bfseries, color=magold!80!black] {Extraction Schema};
        \node[above=0.15cm of json, font=\footnotesize\bfseries, color=black] {Structured Outputs};

    \end{tikzpicture}

    \vspace{-0.25\baselineskip}
    
    \caption{Structured data extraction workflow. The user specifies a typed schema describing the fields to extract. The LLM (or VLM, for image and scanned-document inputs) returns a schema-valid JSON record that maps directly to a row in a data frame.}
    \label{fig:structured_extraction}
\end{figure}
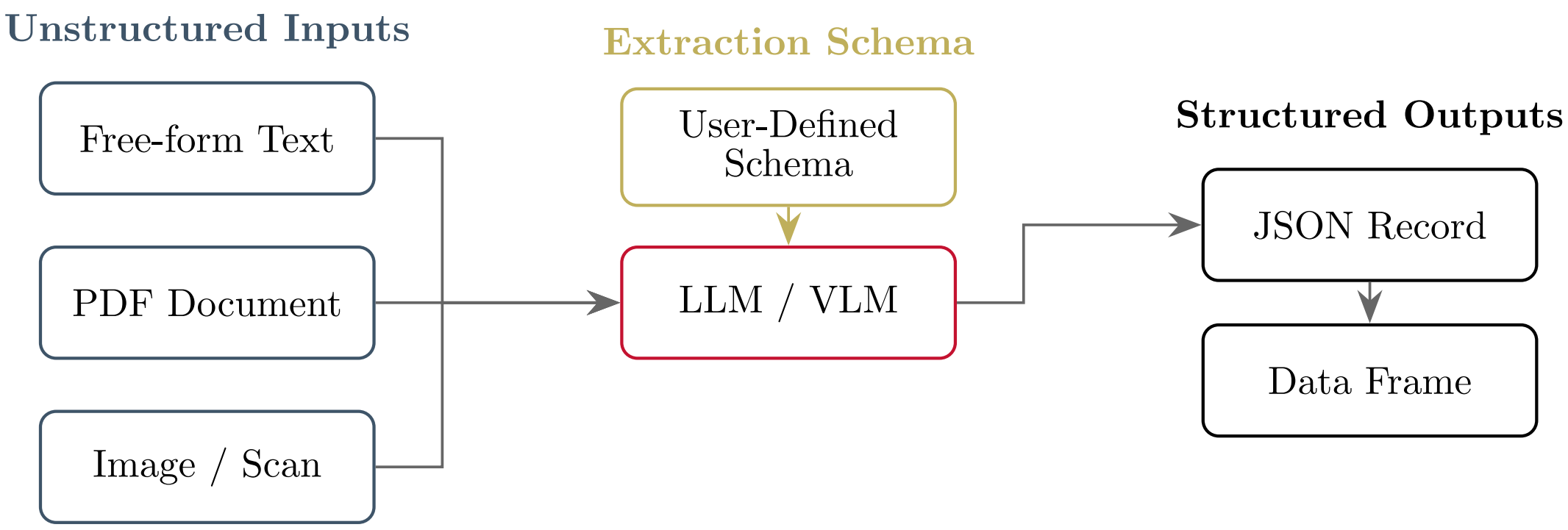

Figure~\ref{fig:structured_extraction} provides an overview of the extraction workflow. We first define the schema, which specifies the variables we want to extract from the document. For each variable, we provide a field name, its data type (e.g., string, integer, or Boolean), and a description written in plain English. We then provide the source document and ask the model to return the extracted information as a JSON record \citep{openai2025structured}. The records can subsequently be combined into a data frame for analysis. Note that this approach can be used with different input formats, including text, PDFs, and images (the latter requiring a VLM).

The field \texttt{description} allows us to be specific about the information we want. For example, when extracting the manufacturer from a recall notice, we may want the full company name, its abbreviation, or both. We can specify this in the description and modify it without changing the extraction code. However, a correctly formatted JSON record does not necessarily mean that the extraction is correct. The model may return all the required fields with the appropriate data types while still extracting incorrect information. This is why we use human reviewers in both our demonstration applications to assess the extracted values against their source documents.

Listings~\ref{lst:ellmer_extract} and~\ref{lst:chatlas_extract} show how to perform structured extraction in R and Python, respectively. Both examples extract two fields from a recall notice: \texttt{manufacturer} and \texttt{defect\_summary}. In R, we use the \texttt{ellmer} package \citep{wickham2025ellmer}, where \texttt{type\_object()} defines the schema and \texttt{chat\_structured()} performs the extraction. In Python, we use \texttt{chatlas} \citep{posit2026chatlas}, which defines the fields using a Pydantic \texttt{BaseModel} and performs the extraction with \texttt{extract\_data()}. The models listed are for illustrative purposes. Model selection should consider accuracy, latency, cost, and privacy needs \citep{singh2026multimodal}.

\begin{figure}[!ht]
\begin{minipage}[t]{0.47\textwidth}
\begin{lstlisting}[basicstyle=\footnotesize\ttfamily,
        language=R,
        commentstyle=\color{miamired},
        backgroundcolor=\color{light-tan!15},
        keywordstyle=\bfseries,
        morekeywords={library,chat_openai,type_object,type_string,chat_structured},
        emphstyle=\bfseries\ttfamily,
        otherkeywords={<-,|>},
        showstringspaces=false,
        breaklines=true,
        caption={Structured extraction in R (\texttt{ellmer}).},
        label={lst:ellmer_extract}]
# --- Load packages ---
library(ellmer)

# --- Initialize provider ---
chat <- chat_openai(
  model = "gpt-6-luna"
)

# --- Define schema ---
schema <- type_object(
  manufacturer = type_string(
    description =
    "Company issuing the recall."
  ),
  defect_summary = type_string(
    description =
    "One-sentence defect summary."
  )
)

# --- Extract structured data ---
result <- chat$chat_structured(
  recall_text,
  type = schema
)
# result$manufacturer
# result$defect_summary
\end{lstlisting}
\end{minipage}%
\hfill
\begin{minipage}[t]{0.47\textwidth}
\begin{lstlisting}[basicstyle=\footnotesize\ttfamily,
        language=Python,
        commentstyle=\color{miamired},
        backgroundcolor=\color{light-tan!15},
        keywords={from, import, class, str, Field,
                  ChatOpenAI, extract_data},
        keywordstyle=\bfseries\ttfamily,
        showstringspaces=false,
        breaklines=true,
        caption={Structured extraction in Python (\texttt{chatlas}).},
        label={lst:chatlas_extract}]
# --- Load packages ---
from chatlas import ChatOpenAI
from pydantic import BaseModel, Field
# --- Initialize provider ---
chat = ChatOpenAI(
    model="gpt-6-luna"
)

# --- Define schema ---
class RecallInfo(BaseModel):
    manufacturer: str = Field(
        description=
        "Company issuing the recall."
    )
    defect_summary: str = Field(
        description=
        "One-sentence defect summary."
    )

# --- Extract structured data ---
result = chat.extract_data(
    recall_text,
    data_model = RecallInfo,
)
# result.manufacturer
# result.defect_summary
\end{lstlisting}
\end{minipage}
\end{figure}

Note that both packages support multiple LLM providers. For example, in \texttt{ellmer}, we can replace \texttt{chat\_openai()} with \texttt{chat\_ollama()} for a locally deployed model, \texttt{chat\_anthropic()} for Claude, or \texttt{chat\_google\_gemini()} for Gemini. The schema and extraction call remain unchanged when switching providers. In the following section, we use this capability to compare two models on the same recall extraction task.

%% file: 04_application_1.tex
\paragraph*{Motivation and practical need.}
Recall notices issued by federal agencies are an example of unstructured data of interest to manufacturing firms. In the U.S., product safety recalls are widespread. For example, in 2025, NHTSA issued 891 recall campaigns affecting more than 29 million vehicles \citep{nhtsa2026annual}. Each recall includes a description of the underlying defect that must be coded, categorized, and tracked. Across all U.S. industries, the \textit{Sedgwick Product Safety and Recall Index} captured 1,634 recalls affecting 941.2 million products in Jan--Jun 2026 \citep{sedgwick2026recalls}. Prior research has demonstrated that specialized natural language processing (NLP) techniques can extract and classify defects from recall notices and related records \citep{francis2023smartxt,uglanov2024nlp}. However, applying such techniques in practice requires both computational resources and expertise that may not exist in many small and medium manufacturing firms (e.g., Tier II and Tier III automotive suppliers). 
Nonetheless, for these firms, recall notices can offer lessons about component failures similar to those they manufacture, informing reviews of potential vulnerabilities in their own products and processes. 
Converting these notices into structured data could help them identify recurring failure modes and prioritize preventive quality-improvement efforts while reducing the time spent manually reviewing and coding reports.
Here, we demonstrate how to leverage LLMs and plain English to address this problem. We provide a free, online application that can be used for extracting any user-defined fields from unstructured text.

\begin{figure}[htb!]
    \centering
    \begin{subfigure}{\linewidth}
        \centering
        \resizebox{\linewidth}{!}{%
        \begin{tikzpicture}[
            >={Stealth[length=3mm]},
            every node/.style={font=\footnotesize},
            textinbox/.style={
                rectangle, rounded corners=4pt, draw=maslate, line width=0.7pt,
                fill=white, minimum height=0.85cm, minimum width=2.2cm,
                text centered, font=\footnotesize
            },
            imginbox/.style={
                rectangle, rounded corners=4pt, draw=mabordeaux, line width=0.7pt,
                fill=white, minimum height=0.85cm, minimum width=2.2cm,
                text centered, font=\footnotesize
            },
            textproc/.style={
                rectangle, rounded corners=4pt, draw=maslate, line width=0.7pt,
                fill=white, minimum height=0.85cm, text centered, font=\footnotesize
            },
            imgproc/.style={
                rectangle, rounded corners=4pt, draw=mabordeaux, line width=0.7pt,
                fill=white, minimum height=0.85cm, text centered, font=\footnotesize
            },
            extractbox/.style={
                rectangle, rounded corners=4pt, draw=magray, line width=0.7pt,
                fill=white, minimum height=0.85cm, text centered, font=\footnotesize
            },
            schemabox/.style={
                rectangle, rounded corners=4pt, draw=magold!80!black, line width=0.7pt,
                fill=white, minimum height=0.85cm, text centered, font=\footnotesize
            },
            outbox/.style={
                rectangle, rounded corners=4pt, draw=black, line width=0.7pt,
                fill=white, minimum height=0.85cm, text centered, font=\footnotesize
            },
            arr/.style={->, line width=0.7pt, color=magray},
            lbl/.style={font=\scriptsize\bfseries},
            groupbox/.style={
                rectangle, rounded corners=6pt, draw=#1, line width=0.5pt,
                dashed, inner sep=6pt
            },
            annot/.style={font=\tiny, color=magray, align=center}
        ]
            % --- Row 1: Text-based inputs (left group) ---
            \node[textinbox] at (0, 0) (paste) {Paste Text};
            \node[textinbox] at (2.8, 0) (txtf) {Text File};
            \node[textinbox] at (5.6, 0) (rpdf) {Readable PDF};
            \node[textinbox] at (8.4, 0) (demo) {Demo Data};
            \node[lbl, color=maslate] at (4.2, 0.85) {Text-Based Inputs};

            % --- Row 1: Image-based inputs (right group) ---
            \node[imginbox] at (11.6, 0) (spdf) {Scanned PDF};
            \node[imginbox] at (14.4, 0) (imgs) {Images};
            \node[lbl, color=mabordeaux] at (13.0, 0.85) {Image-Based Inputs};

            % --- Dashed group boxes ---
            \node[groupbox=maslate, fit=(paste)(rpdf)(demo)] (textgroup) {};
            \node[groupbox=mabordeaux, fit=(spdf)(imgs)] (imggroup) {};

            % --- Row 2: Preprocessing (y=-2.2) ---
            \node[textproc, minimum width=3.2cm] at (4.2, -2.2) (direct) {
                \begin{tabular}{@{}c@{}}Direct Text\\[-2pt]{\scriptsize (character string)}\end{tabular}};
            \node[imgproc, minimum width=4.4cm] at (13.0, -2.2) (magick) {
                \begin{tabular}{@{}c@{}}Image Conversion\\[-2pt]{\scriptsize (\texttt{magick} $\to$ \texttt{content\_image\_file()})}\end{tabular}};

            % --- Single collective arrow from each group ---
            \draw[arr, color=maslate, line width=1pt] (textgroup.south) -- (direct.north);
            \draw[arr, color=mabordeaux, line width=1pt] (imggroup.south) -- (magick.north);

            % --- Annotation: page limit ---
            \node[annot, right=0.15cm of magick.east, anchor=west] {max 5\\pages};

            % --- Row 3: LLM + Schema (y=-4.2) ---
            \node[extractbox, minimum width=3.6cm] at (6.5, -4.2) (llm) {
                \begin{tabular}{@{}c@{}}LLM / VLM\\[-2pt](configurable)\end{tabular}};
            \node[schemabox, minimum width=4.8cm] at (13.0, -4.2) (schema) {
                \begin{tabular}{@{}c@{}}User-Defined Schema\\[-2pt]{\scriptsize labels + plain-English descriptions}\\[-2pt]{\scriptsize $\to$ \texttt{type\_object()} at runtime}\end{tabular}};
            \node[lbl, color=magray] at (5, -3.35) {Extraction Layer}; % (6.5, -3.35)
            \node[lbl, color=magold!80!black] at (13.0, -3.35) {Schema Layer};

            % --- Preprocessing to LLM arrows ---
            \draw[arr] (direct.south) -- ++(0,-0.4) -| (llm.north);
            \draw[arr] (magick.south) -- ++(0,-0.4) -| (llm.north);

            % --- Schema to LLM arrow ---
            \draw[arr, color=magold!80!black] (schema.west) -- (llm.east);

            % --- Row 4: Output (y=-6.2) --- 
            \node[outbox, minimum width=5.0cm] at (6.5, -6.2) (output) {
                \begin{tabular}{@{}c@{}}Structured Table\\[-2pt]{\scriptsize (browser preview + CSV download)}\end{tabular}};
            \node[lbl, color=black] at (5.75, -5.35) {Output};
                % (6.5, -5.35)
            \draw[arr] (llm.south) -- (output.north);

            % --- Annotation: JSON validated ---
            \node[annot, right=0.15cm of output.east, anchor=west] {validated\\JSON $\to$ table};

        \end{tikzpicture}%
        }% end resizebox
        \caption{System architecture of the structured text extraction application, including input handling, preprocessing, runtime schema construction, and LLM/VLM-based extraction.}
        \label{fig:arch_text_extraction_a}
    \end{subfigure}
    
    \vspace{0.75em}

    \begin{subfigure}{\linewidth}
        \centering
          \includegraphics[width=0.98\linewidth]{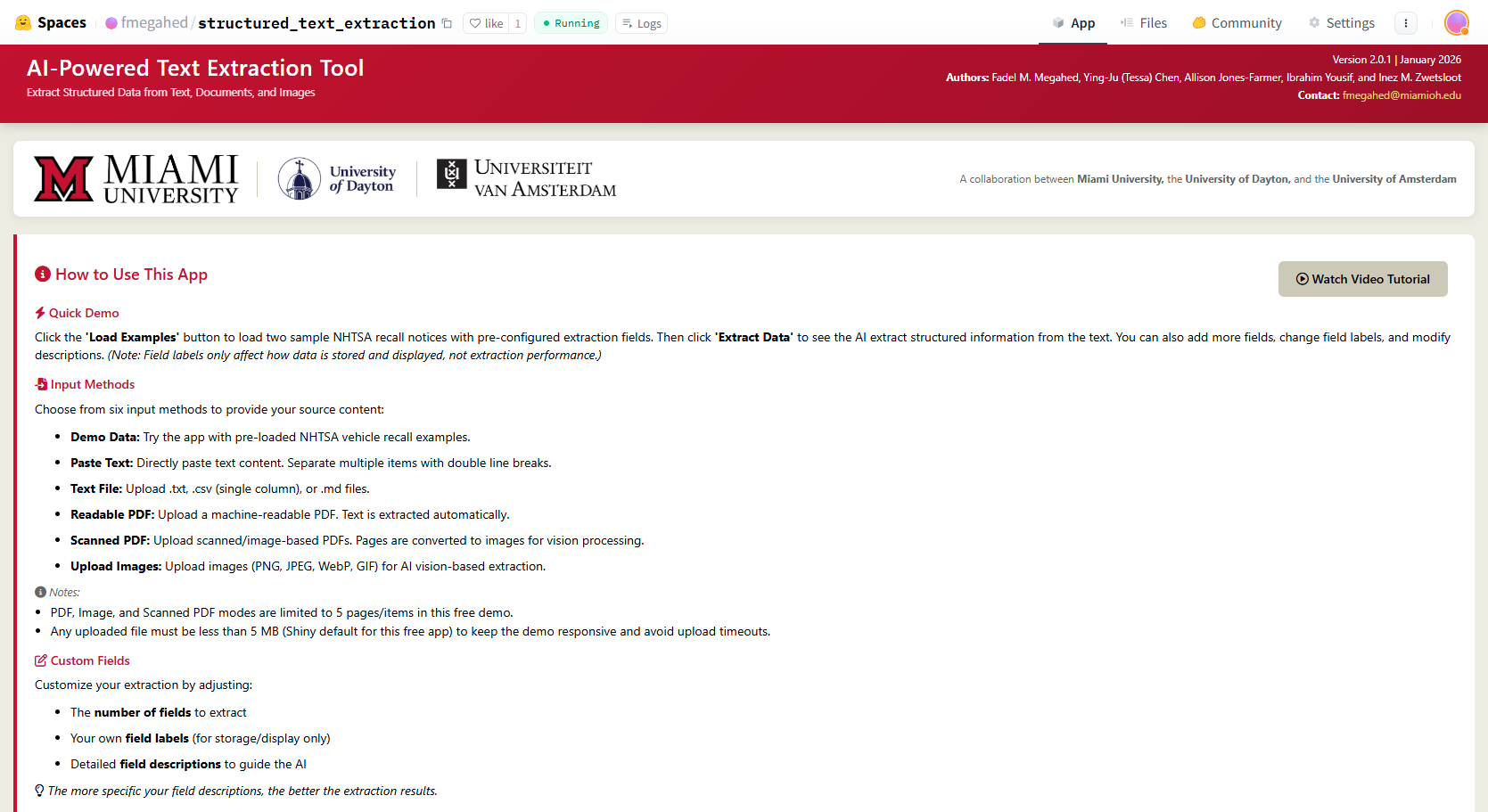}%
        \caption{Hosted Shiny interface (version 2.0.1) showing usage guidance and supported input modes for browser-based structured extraction.}
        \label{fig:arch_text_extraction_b}
    \end{subfigure}

    \caption{Structured text extraction application for quality engineering workflows.}
    \label{fig:arch_text_extraction}
\end{figure}
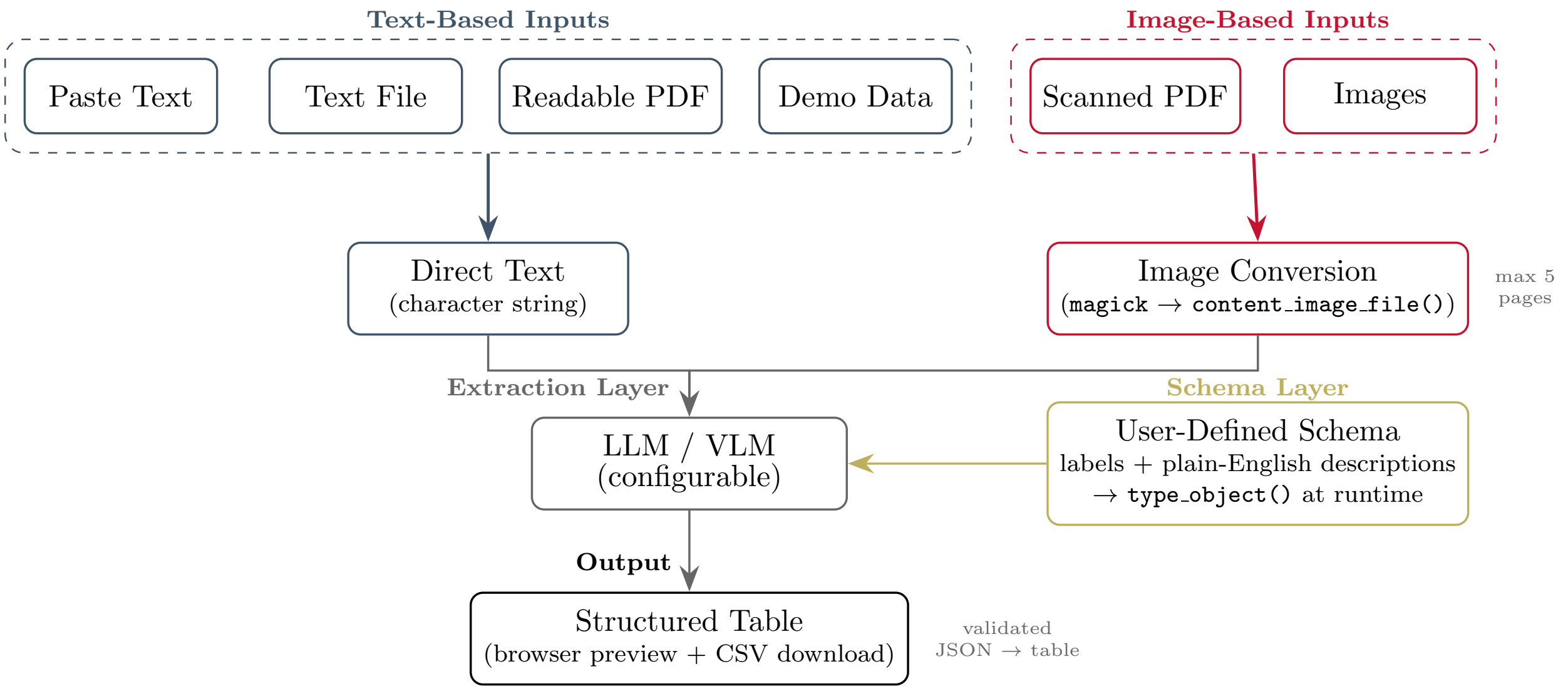
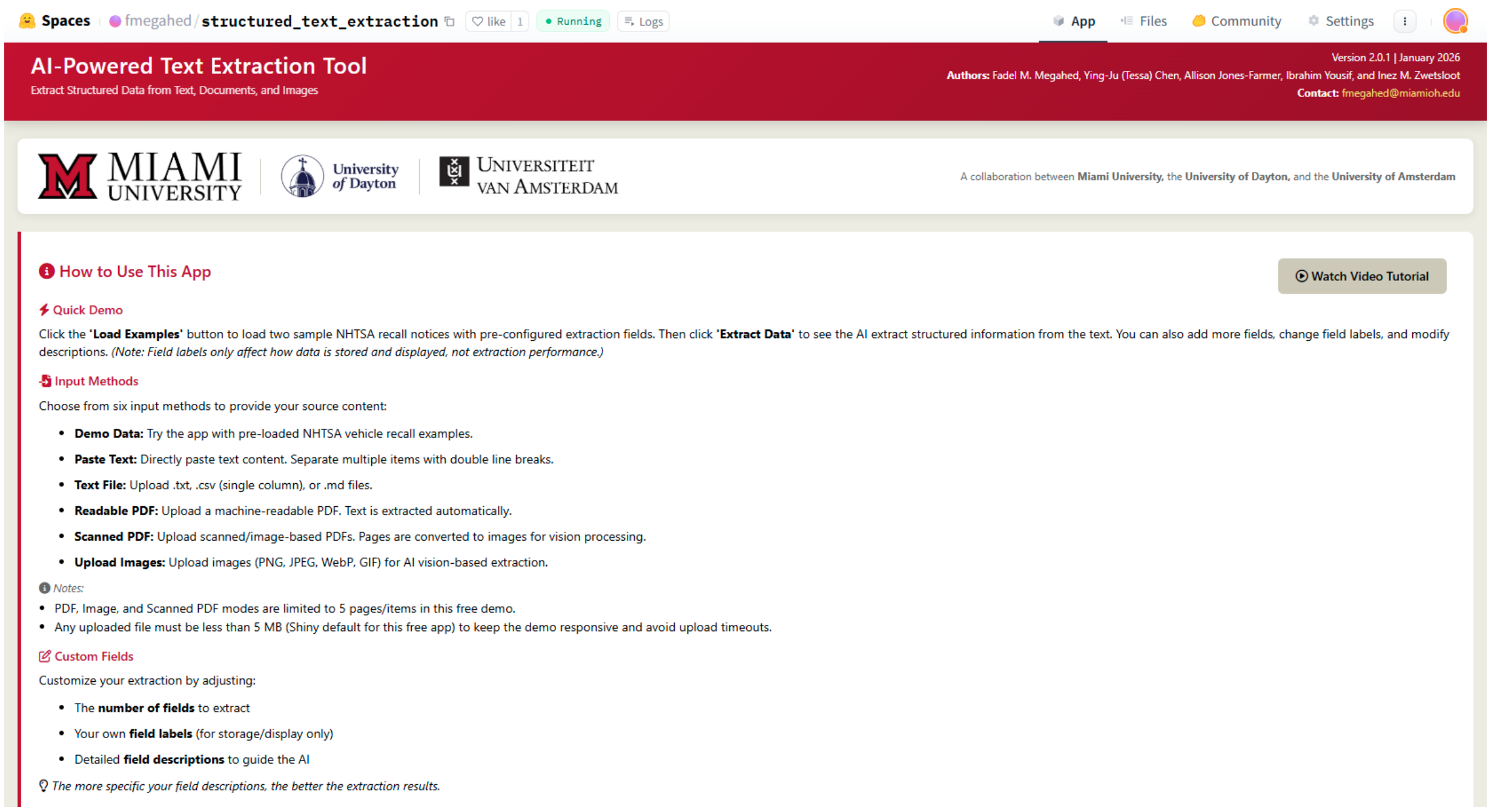

\paragraph*{Application architecture.} To develop our practitioner-facing application, we used the Shiny \texttt{R} package for its development, and we host it at \url{https://huggingface.co/spaces/fmegahed/structured_text_extraction}. In Figure~\ref{fig:arch_text_extraction}, we show the application architecture and the app's user interface as of version 2.0.1, which was publicly available when we started drafting the paper. The later changes are summarized in Section \ref{subsec:improvements}.

The \emph{input layer} supports both text-based and image-based inputs. For text-based inputs, a user can utilize our NHTSA demo sample data to be introduced to text extraction using LLMs. We also enable the user to utilize our Shiny app for their own applications through: (a) pasting text directly into the app's text box, (b) uploading a plain-text file (i.e., a \texttt{.txt} file), or (c) uploading a machine-readable PDF. On the other hand, for image-based inputs, the user can upload a scanned (image-based) PDF for extraction. Furthermore, the user can upload images for text extraction (e.g., reading the labels and values from a bar chart).

The \emph{preprocessing layer} prepares the input content for processing by the AI model. Text inputs and machine-readable PDFs are passed directly as character strings. On the other hand, scanned PDFs and images require an additional step; each page is first converted to an image using the \texttt{magick} R package \citep{ooms2026magick} and then wrapped with \texttt{ellmer's} \texttt{content\_image\_file()} function so that it can be processed by a VLM. Note that we limit the extraction of scanned PDFs to five pages to reduce cost and processing time.

The \emph{schema layer} is where users define the extraction fields. Each field has a label (e.g., \texttt{manufacturer}) and a plain-English description (e.g., ``the company that issued the recall''). Users can add as many fields as needed through the browser interface. These definitions are assembled into an \texttt{ellmer} \texttt{type\_object()} schema at runtime. Because the field descriptions act as natural-language instructions to the model, their specificity directly affects extraction quality. For example, if the user is not happy with how the \texttt{manufacturer's} extraction is performed, they can: (a) specify that they do not want to have both the company name and its short name, indicating which one is needed, and/or (b) provide one- or few-shot learning examples in their prompt to allow the AI model to better grasp the requirements.

The \emph{extraction layer} sends the preprocessed content and schema to the LLM/VLM via an API call. The model returns a JSON record for each input document, which we convert to a structured table where each column corresponds to one extraction field and each row corresponds to one defect/page/etc. For the users' convenience, our Shiny app allows users to inspect the results in their web browser and/or download them as a CSV file.

It is important to note that our Shiny app's usage extends beyond this NHTSA example. Simply put, the user can change the field labels and their descriptions to enable extraction for other applications. This can be done through the user interface for the app in plain English (or any other language of their choice); i.e., the user will not write any R code. 

\paragraph*{Example extraction.}
To make the workflow concrete, consider the two NHTSA recall notices included in the built-in demo. The first notice states (in part): ``Ford Motor Company (Ford) is recalling certain 2021--2022 Bronco vehicles. The rearview camera image may remain displayed on the center stack screen after the vehicle is shifted out of reverse.'' The second notice describes a Honda recall related to a fuel injection ECU software error. Suppose the user defines two extraction fields:

\begin{itemize}[nosep]
    \item \texttt{manufacturer} (description: ``the name of the company recalling vehicles'')
    \item \texttt{defect\_summary} (description: ``summary of the main defect'')
\end{itemize}

\noindent The application sent both notices together with these field definitions to the model and returned the following table:

\begin{table}[htb!]
\centering
\caption{Example extraction from two NHTSA recall notices using the built-in demo. The user defined two fields, and the model populated one row per notice.}
\label{tab:extraction_example}

\vspace{-0.5\baselineskip}

\footnotesize
\begin{tabular}{@{}p{4.25cm} p{12.0cm}@{}}
\toprule
\textbf{manufacturer} & \textbf{defect\_summary} \\
\midrule
Ford Motor Company (Ford) & Rearview camera image may remain displayed after a backing event has ended, causing non-compliance with Federal Motor Vehicle Safety Standard (FMVSS) No. 111, Rear Visibility. \\
Honda (American Honda Motor Co.) & A software error in the fuel injection electronic control unit (FI-ECU) may cause the engine to stall or experience a loss of power. \\
\bottomrule
\end{tabular}
\end{table}

\paragraph*{Deployment and model choice.}
The hosted demo runs on the free CPU tier provided by Hugging Face Spaces. Because the free tier does not provide the hardware needed to run a local VLM/LLM reliably for this application, the demo used OpenAI's \texttt{gpt-5-mini} as the backend when we launched it. For the human evaluation below, we used the newer and less expensive \texttt{gpt-5.4-nano} by changing one input in the code. We will continue to update the demo models as older models are deprecated by OpenAI. Our choice of models provides a practical cost-performance trade-off for structured extraction tasks since our apps are freely and publicly available. Users handling sensitive or proprietary data can run the application on their own infrastructure with a local model, changing only the provider as described in Section~\ref{sec:structured}. The code is publicly available in the Hugging Face repository, and a video tutorial (recorded with version 2.0.1) is available for new users.

\paragraph*{Human evaluation of extraction quality.}
To assess whether the structured outputs accurately represented the source documents, we conducted a blind human evaluation using a sample of the NHTSA recall dataset. Thirty recall descriptions were processed independently using two models: (a) \texttt{gemma4:e2b} (a relatively powerful and small open-weight model that can be deployed without a GPU), and (b) \texttt{gpt-5.4-nano} (one of the cheapest models from OpenAI at the time of the assessment). For each recall, we extracted three fields: manufacturer, vehicle models, and model years. We chose these fields since they can be checked directly against the recall description without subjective interpretation.

The resulting extractions were presented in randomized order without identifying the model that produced them. To administer the evaluation, we built a web-based review application that provided each reviewer with a personalized link. Our application simplified the review process by: (a) having each screen show one recall description alongside one LLM's extraction for the three fields; and (b) making the rubric definitions available on every screen. Four reviewers (pooled from students and research assistants) evaluated each field against the recall description using a five-point correctness scale ranging from 1 (Incorrect) to 5 (Correct), with intermediate categories distinguishing partially correct extractions and those containing minor errors or omissions. This rubric is presented in full and discussed further in Section~\ref{sec:app2}. Reviewers were instructed to judge each extraction only against the provided recall text. The app's source code and a short overview video are available at the GitHub repository provided in our ``Reproducibility materials'' section.

Across the 720 field-level ratings, 710 (98.6\%) received a rating of 5 (Correct). Table~\ref{tab:recall_eval} shows that both models performed consistently well across the three fields.

\begin{table}[htbp]
\centering
\caption{Human evaluation of structured fields extracted from 30 NHTSA recall descriptions. Entries report the percentage of human ratings equal to 5 (Correct).}
\label{tab:recall_eval}
\begin{tabular}{lrrrr}
\toprule
Model & Manufacturer & Vehicle Models & Model Years & Overall \\
\midrule
\texttt{gemma4:e2b}    & 100.0\% & 97.5\% & 98.3\% & 98.6\% \\
\texttt{gpt-5.4-nano}  & 100.0\% & 96.7\% & 99.2\% & 98.6\% \\
\midrule
Overall                 & 100.0\% & 97.1\% & 98.8\% & 98.6\% \\
\bottomrule
\end{tabular}
\end{table}

The four human reviewers also showed strong agreement in their evaluations. They assigned identical ratings to 174 of the 180 model--document--field combinations (96.7\%). Of the six combinations with disagreement, five involved ratings of 4 versus 5. The remaining case concerned extracting vehicle models from a recall description. The source indicated that \texttt{PACCAR} was recalling certain \texttt{2021-2025 Peterbilt} and \texttt{Kenworth} vehicles but referred readers to the recall report for a complete list of models. The \texttt{gpt-5.4-nano} model did not extract any vehicle models, which three reviewers rated as incorrect (1), while one rated it as mostly correct (4). The disagreement appears to stem from how reviewers interpreted the field being evaluated. Although \texttt{Peterbilt} and \texttt{Kenworth} were explicitly mentioned in the source, they are vehicle brands rather than specific models. Thus, the model's decision not to extract them was defensible given the information provided. This case highlights that even when evaluating fields that can be verified directly against a source, differences in how those fields are defined and interpreted can lead to disagreement among human reviewers.

%% file: 05_application_2.tex
\paragraph*{Motivation and practical need.} We presume that readers may have used ChatGPT (or a similar tool) to summarize a PDF or to find relevant literature on a quality engineering topic. While the (technical) summary generated is often useful, by default, it tends to be quite verbose, spread across multiple paragraphs, and may omit aspects important to the reader if the LLM was not explicitly prompted to capture them. Furthermore, searches (which have improved considerably since the earlier ChatGPT days) tend to be one-time tasks that users need to repeat periodically to identify new papers. Note that comparisons across sessions/papers are also difficult since the responses do not follow a common format. To address these limitations, we developed \texttt{QE ArXiv Watch}. It performs a daily check of arXiv for new papers in quality engineering, reads the full text of each paper, and extracts more than 20 predefined fields from each paper. It then presents the results in a web-based interface through multiple avenues: (a) filterable tables, (b) visualizations, (c) a page for each paper, and (d) aggregations by discipline/author.  QE ArXiv Watch is freely available at: \url{https://huggingface.co/spaces/fmegahed/arxiv_control_charts}.

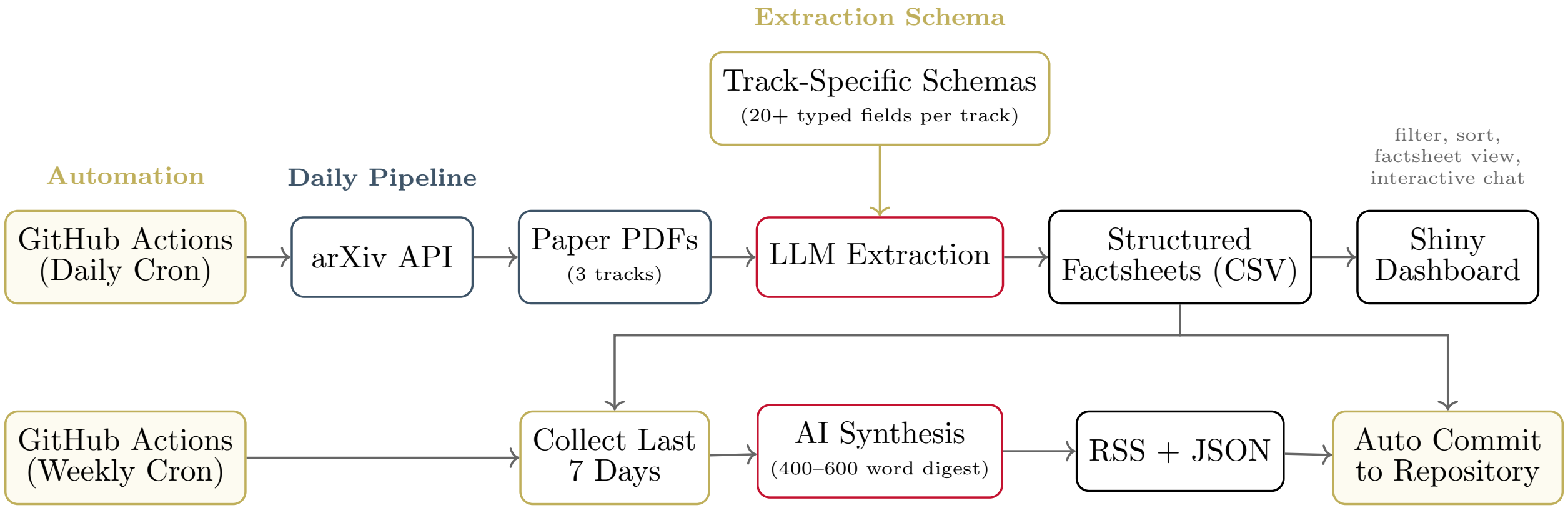
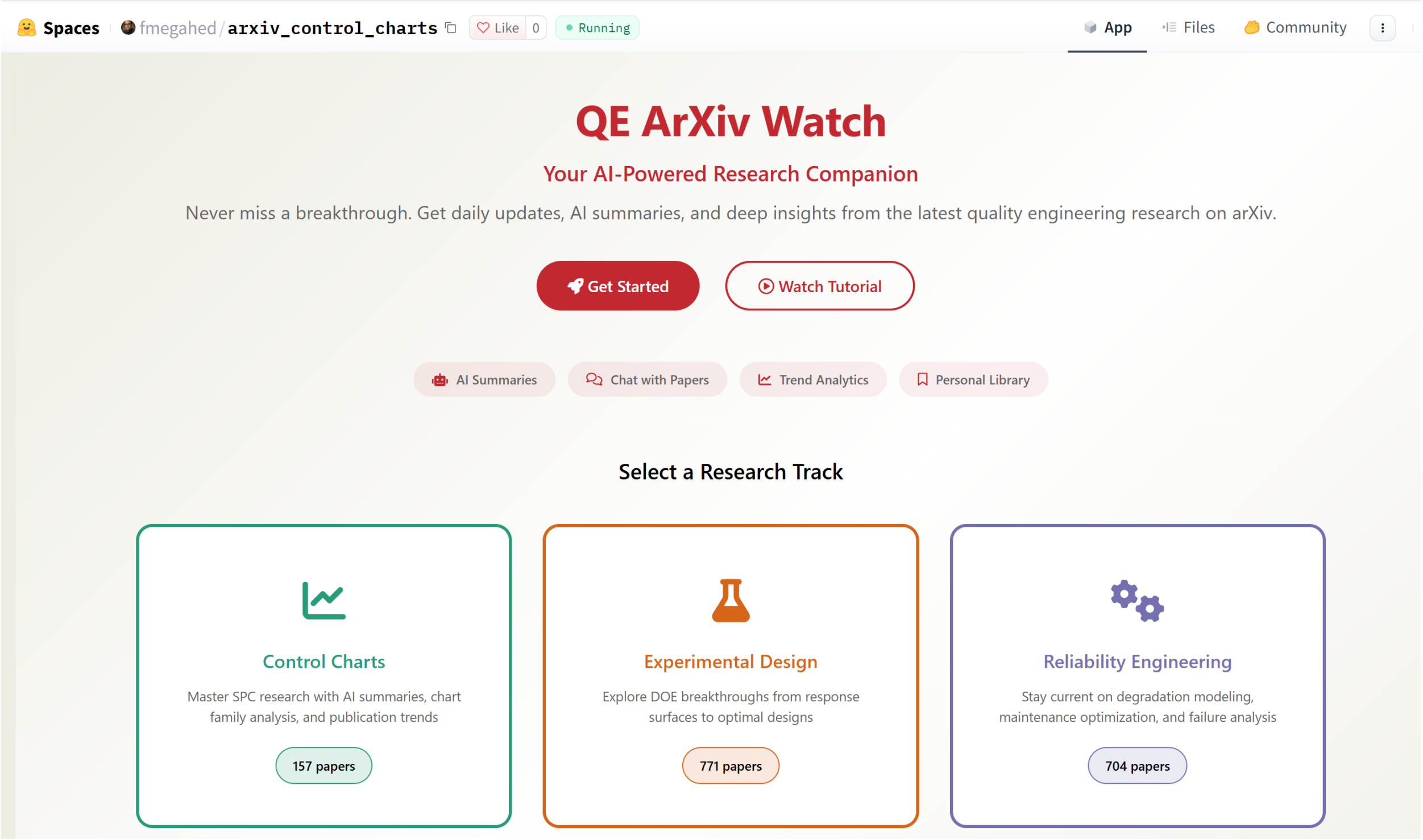
\begin{figure}[htb!]
    \centering

    % -------- Panel (a): Architecture --------
    \begin{subfigure}{\linewidth}
        \centering
        \resizebox{\linewidth}{!}{%
        \begin{tikzpicture}[
            every node/.style={font=\footnotesize},
            stepnode/.style={
                rectangle, rounded corners=4pt, draw=maslate, line width=0.7pt,
                fill=white, minimum height=0.9cm, minimum width=2.05cm,
                text centered, font=\footnotesize
            },
            procnode/.style={
                rectangle, rounded corners=4pt, draw=mabordeaux, line width=0.7pt,
                fill=white, minimum height=0.9cm, minimum width=2.05cm,
                text centered, font=\footnotesize
            },
            outnode/.style={
                rectangle, rounded corners=4pt, draw=black, line width=0.7pt,
                fill=white, minimum height=0.9cm, minimum width=2.05cm,
                text centered, font=\footnotesize
            },
            schednode/.style={
                rectangle, rounded corners=4pt, draw=magold!80!black, line width=0.7pt,
                fill=white, minimum height=0.9cm, minimum width=2.05cm,
                text centered, font=\footnotesize
            },
            ghanode/.style={
                rectangle, rounded corners=4pt, draw=magold!80!black, line width=0.7pt,
                fill=magold!10, minimum height=0.9cm, minimum width=2.05cm,
                text centered, font=\footnotesize
            },
            dashnode/.style={
                rectangle, rounded corners=4pt, draw=black, line width=0.7pt,
                fill=white, minimum height=0.9cm, minimum width=2.05cm,
                text centered, font=\footnotesize
            },
            arr/.style={->, line width=0.7pt, color=magray},
            annot/.style={font=\tiny, color=magray, align=center}
        ]
            % --- GitHub Actions trigger ---
            \node[ghanode] (gha) {
                \begin{tabular}{@{}c@{}}GitHub Actions\\[-2pt](Daily Cron)\end{tabular}};

            % --- Daily pipeline ---
            \node[stepnode, right=0.5cm of gha] (arxiv) {arXiv API};
            \node[stepnode, right=0.5cm of arxiv] (meta) {
                \begin{tabular}{@{}c@{}}Paper PDFs\\[-2pt]{\tiny (3 tracks)}\end{tabular}};
            \node[procnode, right=0.5cm of meta] (extract) {
                \begin{tabular}{@{}c@{}}LLM Extraction \end{tabular}};
            \node[outnode, right=0.5cm of extract] (facts) {
                \begin{tabular}{@{}c@{}}Structured\\[-2pt]Factsheets (CSV)\end{tabular}};
            \node[dashnode, right=0.5cm of facts] (dash) {
                \begin{tabular}{@{}c@{}}Shiny\\[-2pt]Dashboard\end{tabular}};

            \draw[arr] (gha) -- (arxiv);
            \draw[arr] (arxiv) -- (meta);
            \draw[arr] (meta) -- (extract);
            \draw[arr] (extract) -- (facts);
            \draw[arr] (facts) -- (dash);

            % --- Schema feeding into extraction ---
            \node[schednode, above=0.8cm of extract, minimum width=3.4cm] (schema) {
                \begin{tabular}{@{}c@{}}Track-Specific Schemas\\[-2pt]{\tiny (20+ typed fields per track)}\end{tabular}};
            \draw[arr, color=magold!80!black] (schema.south) -- (extract.north);

            % --- Weekly pipeline ---
            \node[ghanode, below=1.2cm of gha] (ghaw) {
                \begin{tabular}{@{}c@{}}GitHub Actions\\[-2pt](Weekly Cron)\end{tabular}};
            \node[schednode, below=1.2cm of meta] (weekly) {
                \begin{tabular}{@{}c@{}}Collect Last\\[-2pt]7 Days\end{tabular}};
            \node[procnode, below=1.2cm of extract] (synth) {
                \begin{tabular}{@{}c@{}}AI Synthesis\\[-2pt]{\tiny (400--600 word digest)}\end{tabular}};
            \node[outnode, below=1.2cm of facts] (rss) {RSS + JSON};

            \draw[arr] (ghaw) -- (weekly);
            \draw[arr] (facts.south) -- ++(0,-0.35) -| (weekly.north);
            \draw[arr] (weekly) -- (synth);
            \draw[arr] (synth) -- (rss);

            % --- Git commit arrow ---
            \node[ghanode, below=1.2cm of dash, minimum width=2.6cm] (commit) {
                \begin{tabular}{@{}c@{}}Auto Commit\\[-2pt]to Repository\end{tabular}};
            \draw[arr] (facts.south) -- ++(0,-0.35) -| (commit.north);
            \draw[arr] (rss) -- (commit);

            % --- Dashboard features annotation ---
            \node[annot, above=0.15cm of dash, text width=2.2cm] {filter, sort,\\factsheet view,\\interactive chat};

            % --- Labels ---
            \node[above=0.15cm of gha, font=\scriptsize\bfseries, color=magold!80!black] {Automation};
            \node[above=0.15cm of arxiv, font=\scriptsize\bfseries, color=maslate] {Daily Pipeline};
            \node[above=0.15cm of schema, font=\scriptsize\bfseries, color=magold!80!black] {Extraction Schema};

        \end{tikzpicture}%
        }
        \caption{System architecture with GitHub Actions automation for daily arXiv ingestion, LLM-based structured extraction, weekly digest generation, and repository updates.}
        \label{fig:arch_arxiv_a}
    \end{subfigure}

    \vspace{1em}

    % -------- Panel (b): Dashboard screenshot --------
    \begin{subfigure}{\linewidth}
        \centering
        \includegraphics[width=\linewidth]{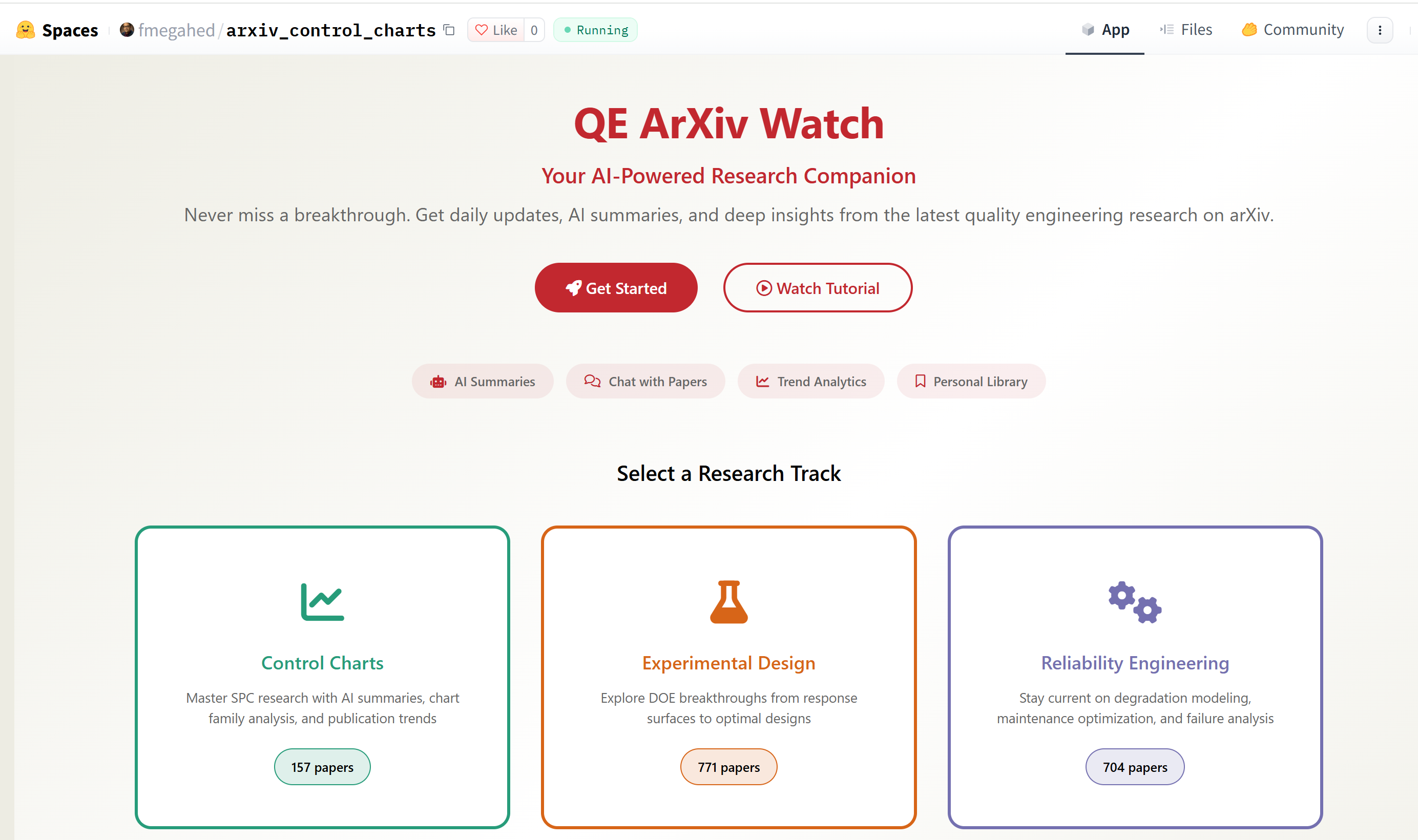}
        \caption{The landing page showing the three research tracks, paper counts by track, and summary indicators used to access the dashboard's structured factsheets and related tools.}
        \label{fig:arch_arxiv_b}
    \end{subfigure}

    \caption{Architecture and landing page of QE ArXiv Watch.}
    \label{fig:arch_arxiv}
\end{figure}

\paragraph*{Application architecture.}
The application has three components: (a) a daily extraction pipeline, (b) a weekly digest pipeline, and (c) an interactive dashboard. The two pipelines run on a schedule via \texttt{GitHub Actions}, which ensures that \texttt{QE ArXiv Watch} updates automatically. In Figures~\ref{fig:arch_arxiv_a}--\ref{fig:arch_arxiv_b}, we provide a summary of \texttt{QE ArXiv's} architecture and show the landing page with the number of papers processed as of October 08, 2026. The landing page's screenshot is based on version 3.3.0, the version whose factsheets were evaluated by a sample of the arXiv paper authors. 

The \emph{daily extraction pipeline} is an R script triggered by a GitHub Actions cron (i.e., scheduled) job each morning. It queries the arXiv API for new papers across three research tracks: (a) statistical process control (SPC; named statistical process monitoring in the current version), (b) experimental design (DOE), and (c) reliability engineering. In the evaluated version, the SPC track searched for papers containing terms such as ``control chart'' in the title or abstract. The DOE track searched for terms such as ``experimental design,'' ``response surface,'' and ``supersaturated design.'' Finally, the reliability track searched for terms such as ``reliability,'' ``degradation,'' ``maintenance,'' and ``remaining useful life,'' with restrictions to relevant statistics categories. We later revised these searches (Section~\ref{subsec:improvements}). When new papers are identified, the script reads each PDF directly from arXiv (without persisting a local copy) using \texttt{ellmer's} \texttt{content\_pdf\_url()} function and sends it to an LLM together with a track-specific extraction schema. The resulting structured factsheets are appended to CSV files redundantly stored in both our GitHub and HuggingFace repositories.

The \emph{weekly digest pipeline} is a second GitHub Actions job that runs every Monday. It collects all factsheets created during the previous seven days and sends them to the LLM with instructions to prepare a 400--600 word summary of the new research (published within the past seven days). The goal is to give researchers a brief overview of recent papers without requiring them to read each factsheet. The application publishes the summary as both an RSS feed (which readers can subscribe to) and a JSON file (easily read in R/Python).

The \emph{interactive dashboard} is a Shiny application hosted on Hugging Face Spaces. It reads the CSV files produced by the daily pipeline and presents three track-specific pages plus a landing page with summary statistics. On each track page, users can filter papers by extracted fields (e.g., simulation-based EWMA papers), sort by date or other extracted dimensions, and view a paper's full factsheet in a pop-up panel. The dashboard also includes an interactive chat feature, allowing users to ask follow-up questions about a paper.

\paragraph*{Extraction schemas.}
The extraction schemas in this application are more detailed than those in the practitioner-oriented example because they must capture the dimensions researchers use to classify papers.
We selected a more capable model (\texttt{gpt-5.2} in the evaluated version), as the task requires reading full research papers in PDF form and populating many fields with a consistent structure. Several fields are also multi-label enumerations (for example, a single paper may belong to multiple chart families or report multiple performance metrics), which increases extraction complexity. The SPC schema alone contains more than twenty typed fields organized into five groups (Table~\ref{tab:spc_schema}). Similar schemas are used for the DOE and reliability tracks, with domain-appropriate fields (e.g., design type, design objective, optimality criterion, and number of factors for DOE; reliability topic, modeling approach, data type, and maintenance policy for reliability). We have since changed both the schema and the models (Section \ref{subsec:improvements}).

\begin{table}[htb!]
\centering

\caption{Field groups in the SPC extraction schema of the evaluated version.}
\label{tab:spc_schema}

\vspace{-0.5\baselineskip}

\footnotesize
\begin{tabular}{@{}p{3.2cm} p{12.3cm}@{}}
\toprule
\textbf{Group} & \textbf{Fields} \\
\midrule
Classification & Chart family (e.g., Shewhart, CUSUM, EWMA); charting statistic; Phase~I versus Phase~II designation; application domain \\
Data assumptions & Whether the paper assumes normality, addresses autocorrelation, or handles missing data \\
Evaluation & Evaluation type (simulation, real data, theoretical); performance metrics (e.g., ARL or standard deviation of run length); sample-size requirements \\
Software & Whether code was used; platform (R, Python, MATLAB, SAS); whether code is publicly available; where it is hosted \\
Content & Concise summary; key equations; key results; stated limitations; unstated limitations identified by the model; stated future work; future work suggested by the model \\
\bottomrule
\end{tabular}
\end{table}

\paragraph*{Example extraction.}
To illustrate the output, consider the arXiv paper \texttt{2512.23602v1} by \citet{burger2025distributionfree}, titled ``Distribution-Free Process Monitoring with Conformal Prediction'' (submitted December~29, 2025). The daily pipeline downloaded the PDF, sent it to \texttt{gpt-5.2} with the SPC schema, and produced the factsheet shown in Table~\ref{tab:arxiv_extraction_example}.

\begin{table}[htb!]
\centering

\caption{A subset of the extracted fields for arXiv paper \texttt{2512.23602v1} using the SPC schema (evaluated version). Pipe-delimited values indicate multi-label fields. The values \texttt{Both} and \texttt{Other} in this table are examples of the problems that the authors reported, which are addressed as explained in Section \ref{subsec:improvements}.}
\label{tab:arxiv_extraction_example}

\vspace{-0.5\baselineskip}

\footnotesize
\begin{tabular}{@{}p{3.5cm} p{12cm}@{}}
\toprule
\textbf{Field} & \textbf{Extracted Value} \\
\midrule
\texttt{chart\_family} & Univariate \textbar{} Multivariate \textbar{} Nonparametric \textbar{} High-dimensional \textbar{} Other\\
\texttt{chart\_statistic} & Shewhart \textbar{} Machine learning-based \textbar{} Other \\
\texttt{phase} & Both \\
\texttt{assumes\_normality} & FALSE \\
\texttt{evaluation\_type} & Simulation study \textbar{} Other \\
\texttt{performance\_metrics} & False alarm rate \textbar{} Detection probability \textbar{} Other \\
\texttt{code\_used} & TRUE \\
\texttt{code\_availability} & Public repository (GitHub) \\
\texttt{code\_url} & \url{https://github.com/christopherburger/ConformalSPC} \\
\texttt{summary} & The paper proposes a hybrid Statistical Process Control (SPC) framework that integrates Conformal Prediction (CP) to obtain distribution-free, model-agnostic monitoring guarantees under exchangeability rather than normality. It introduces (i) Conformal-Enhanced Control Charts that replace parametric control limits with a conformal threshold derived from calibration nonconformity scores, and optionally plot conformal prediction intervals over time to visualize uncertainty and trigger ``uncertainty spike'' signals. It also proposes (ii) Conformal-Enhanced Process Monitoring that reframes multivariate SPC as online anomaly detection and monitors a single conformal p-value over time via a p-value chart with a control limit at significance level $\alpha$. Nonconformity scores are defined using robust center measures (e.g., median-based absolute deviations) or model-based residuals (possibly normalized by an estimated local scale), allowing use with arbitrary predictive/anomaly-detection models (e.g., SVMs, autoencoders). The paper provides simulated chart comparisons for normal and exponential data and argues improved robustness and sensitivity relative to traditional Shewhart-style charts while maintaining interpretability. \\
\bottomrule
\end{tabular}
\end{table}

\noindent The \texttt{assumes\_normality} field is \texttt{FALSE}, which is consistent with the paper's emphasis on distribution-free monitoring via conformal prediction. Once extracted, this factsheet becomes one row in the dashboard's SPC table. A researcher can then filter the dashboard to identify all papers that do not assume normality, or all papers with publicly available code, thereby turning part of the literature review process into a structured query task.

\paragraph*{Current scale and aggregate insights.}
As of October 08, 2026, the evaluated version contained structured factsheets for 1,632 unique papers: 157 in the SPC track (spanning submissions from 2008 to Oct 2026), 771 in DOE, and 704 in reliability.

Aggregate analyses that would be difficult to produce manually are available directly from the structured factsheets and dashboard filters. For example, across the 157 SPC papers, the most common chart families are univariate (93 papers) and multivariate (49), followed by nonparametric (36), profile monitoring (17), and high-dimensional methods (16). Only 58 of the 157 papers are coded as assuming normality, suggesting sustained activity in distribution-free and robustness-oriented methods. R is the most commonly identified software platform (55 papers), followed by Python (12) and MATLAB (9). The code-availability pattern is also notable: only 12 papers provide code in a public repository, and only 10 provide code through a package registry (CRAN or PyPI). By contrast, 111 papers are coded as using software but not providing publicly accessible code.

\paragraph*{Quality checks and author evaluation.}
To assess the quality of the AI-generated factsheets, we invited authors to evaluate the extracted information for their papers. We randomly selected 30 papers from each track, excluded those coauthored by members of our team, and invited one author per remaining paper (85 invitations, of which 81 were delivered). Authors could evaluate any of their papers included in the dashboard. In total, 21 authors evaluated 45 factsheets: 26 SPC, 14 reliability, and 5 DOE. 

For each factsheet, authors rated eight fields: four common to all tracks (AI Summary, Key Results, Key Equations, and Unstated Future Work) and four track-specific classification fields. We used the same five-point correctness scale introduced in the first application, consistent with prior work on rating scales and expert evaluation of LLM outputs \citep{allen2007likert,megahed2024introducing}. For Unstated Future Work, we used a parallel five-point reasonableness scale, since this field represents an AI-generated suggestion rather than information extracted directly from the paper. Table~\ref{tab:evaluation_scales} provides the definitions for both scales. Authors could also provide comments, which we used to identify extraction problems and refine the prompts and schemas.

\begin{table}[htbp!]
\centering
\caption{Scales to assess structured extraction outputs and future-work reasonableness.}
\label{tab:evaluation_scales}

\vspace{-0.5\baselineskip}

\footnotesize

\setlength{\tabcolsep}{4pt}

\begin{tabularx}{\textwidth}{@{}c
    >{\raggedright\arraybackslash}p{1.5cm}
    >{\raggedright\arraybackslash}X
    >{\raggedright\arraybackslash}p{2.05cm}
    >{\raggedright\arraybackslash}X@{}}
\toprule
& \multicolumn{2}{c}{\textbf{Correctness}} 
& \multicolumn{2}{c}{\textbf{Reasonableness}} \\
\cmidrule(lr){2-3}\cmidrule(l){4-5}
\textbf{Score} 
& \textbf{Anchor} 
& \textbf{Definition} 
& \textbf{Anchor} 
& \textbf{Definition} \\
\midrule
1 
& Incorrect 
& The response is wrong, unsupported by the paper, or clearly misrepresents the paper's content. 
& Unreasonable 
& The suggested future work is implausible, disconnected from the paper, or not meaningfully motivated by its content. \\

2 
& Mostly Incorrect 
& The response contains a small amount of correct information, but substantial errors, misclassifications, or important omissions remain. 
& Mostly Unreasonable 
& The suggestion has limited connection to the paper and is only weakly justified by its methods, findings, or discussion. \\

3 
& Partially Correct 
& The response is a mix of correct and incorrect information, or it captures only part of what is supported by the paper. 
& Somewhat Reasonable 
& The suggestion is partially plausible and somewhat related to the paper, but the justification is incomplete or only moderately convincing. \\

4 
& Mostly Correct 
& The response is largely correct and supported by the paper, with only minor errors, omissions, or imprecision. 
& Mostly Reasonable 
& The suggestion is plausible, relevant to the paper, and reasonably supported by its content, with only minor weaknesses in justification. \\

5 
& Correct 
& The response is correct, supported by the paper, and does not contain meaningful errors or omissions. 
& Reasonable 
& The suggested future work is plausible, well aligned with the paper, and strongly justified by its content. \\
\bottomrule
\end{tabularx}
\end{table}

\begin{figure}[htb!]
    \centering
    \includegraphics[width=0.9\linewidth]{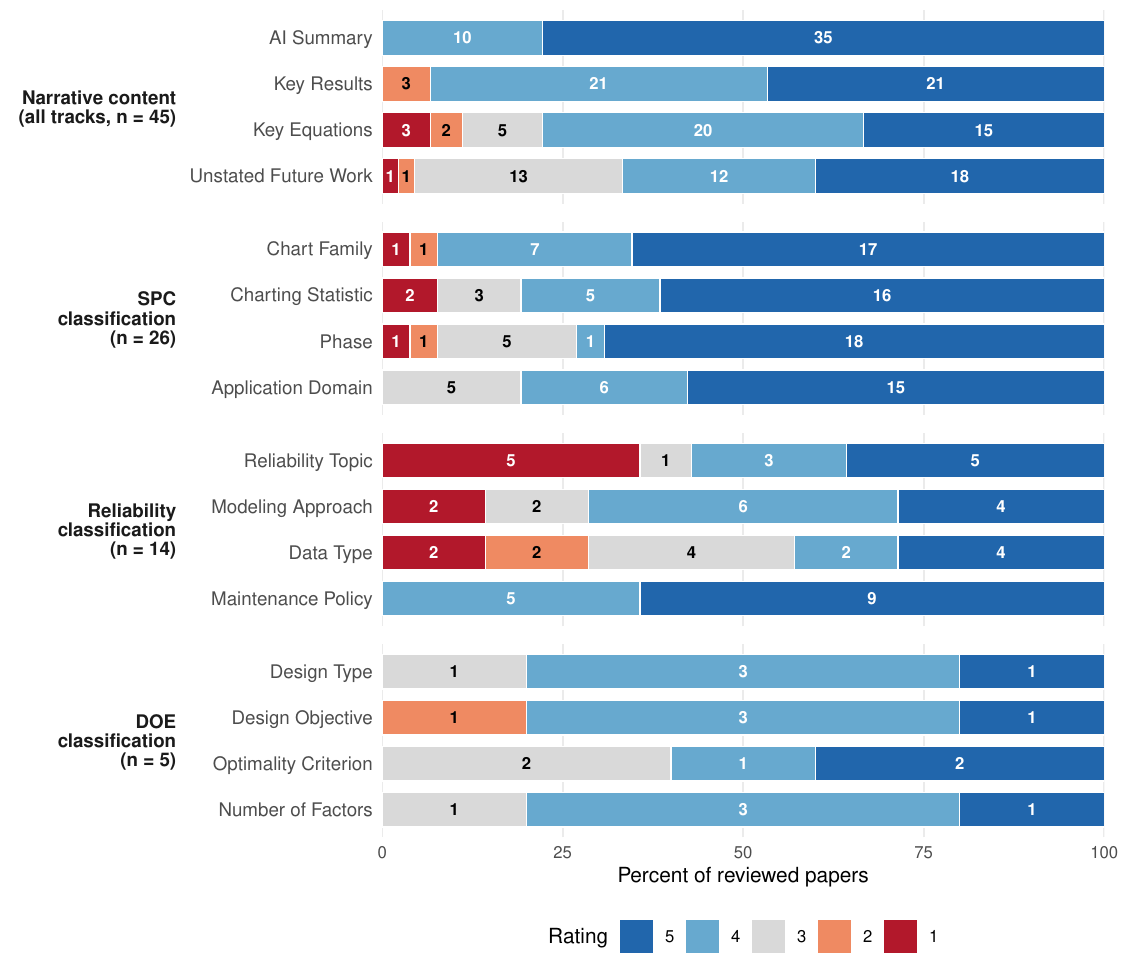}

    \vspace{-0.5\baselineskip}
    
    \caption{Distribution of author ratings for each AI-extracted factsheet field. Narrative fields are common to all tracks; classification fields are track-specific. Numbers inside bars are counts of papers; ratings range from 1 (Incorrect/Unreasonable) to 5 (Correct/Reasonable).}
    \label{fig:arXiv_likert_ratings}
\end{figure}

Figure~\ref{fig:arXiv_likert_ratings} summarizes the author evaluations. The results suggest that the factsheets provide a useful first-pass extraction, although accuracy varies across fields and tracks.

Among the four narrative fields, AI Summary received the highest ratings, with all evaluations at 4 or 5, followed by Key Results (93.3\%), Key Equations (77.8\%), and Unstated Future Work (66.7\%). Several of the lower ratings for Key Equations were associated with mathematical notation that did not render correctly (12 of the 45 factsheets received comments about rendering problems, including all three cases rated 1 for this field). For Unstated Future Work, several authors considered the suggestions reasonable but incremental.

The classification results varied by track. For the SPC track, 80.8\%--92.3\% of ratings for chart family, charting statistic, and application domain were 4 or 5. Among the SPC categorizations, the phase classification labeling had the most room for improvement (73.1\% of ratings were in the 4--5 range). The reviewers noted that their papers addressed only Phase II monitoring or used self-starting charts for which the Phase I/Phase II distinction was not applicable. Reliability had the lowest ratings, particularly for Reliability Topic (57.1\%) and Data Type (42.9\%). Some of these ratings reflected papers retrieved by the reliability search whose main topics (e.g., explainability or active learning) did not fit the schema's categories. Four ratings of 1 were also associated with missing values displayed as \texttt{nan}. We do not comment on the DOE results since only five papers were evaluated, which makes the sample size too small to draw meaningful conclusions. We used these ratings and comments to revise the app, as described in Section \ref{subsec:improvements}.

%% file: 06_discussion.tex
\subsection{Lessons from the Applications}\label{subsec:lessons}

One observation from our two applications is that extraction quality depends on the task and the fields being extracted. In the recall application, two relatively small models performed well when extracting factual information such as manufacturer, vehicle models, and model years. In \texttt{QE ArXiv Watch}, the AI-generated summaries generally received high ratings, whereas some of the classification fields received considerably lower ratings. Although the two evaluations involved different tasks and cannot be directly compared, they suggest that evaluating an extraction application using a single overall accuracy measure may be misleading. Instead, we recommend examining the quality of individual fields, particularly those that will be used in subsequent analyses.

Another lesson concerns the sources of extraction errors. In both applications, human reviewers identified issues related to how fields were defined/interpreted. In \texttt{QE ArXiv Watch}, some of the lower ratings were also associated with problems in displaying mathematical equations and handling missing values. Note that these issues cannot necessarily be resolved by switching to a more powerful model. Instead, they may require changes to the schema, extraction prompts, or the application itself. This is one reason why we encourage users to review a sample of the extracted records and use their findings to improve the extraction workflow. We describe the changes we made in response to the identified issues in the next subsection.

\subsection{Improvements and Changes Stemming from Author Review}\label{subsec:improvements}

The author evaluations of \texttt{QE ArXiv Watch} helped us identify several problems with different causes. For example, equations appearing as raw \LaTeX{} were a display issue that required changes to how the application stores and renders mathematical expressions. Other problems were related to the search terms and extraction schemas. The original reliability search was too broad, retrieving papers containing ``reliable'' that were not related to reliability engineering. To improve the searches, we examined the terminology used in 13,780 articles published in eight ``quality and reliability'' journals. We then tested the search terms against arXiv papers and retained those that mostly retrieved relevant papers. We also added a screening step that evaluates each paper's title and abstract to determine whether it belongs to the corresponding research track. For the extraction schemas, we revised the classification fields to require a single primary category, provided clearer definitions, and added a ``not applicable'' option where needed. This addressed problems such as assigning too many categories to a paper or incorrectly classifying self-starting control charts as Phase~I or Phase~II. We also corrected several issues in the recall application, including a problem where one failed document could cause the results from other documents to be discarded.

Another change was to use multiple models for extracting the classification fields. The revised approach uses a small LLM (currently \texttt{gpt-6-luna}) to read each paper and complete its factsheet. A second model (\texttt{Jev}), from a different provider, independently evaluates the main classification fields \citep{typesafe2026jev}. Unlike the LLM, this second model selects from predefined options and returns a probability for its selection without generating text. When the two models agree, we retain their classification. Otherwise, we ask a larger LLM (currently \texttt{gpt-6.1-sol}) to read the paper and resolve the disagreement. This is similar to having a second inspector evaluate the same part, with a third inspector resolving disagreements. We also indicate in each factsheet whether the classification was agreed upon or required the larger model. In our evaluation of 125 papers, the two models agreed on 82\% of the single-valued classifications, and the larger model was needed for 34\% of the in-scope papers. The resulting cost per in-scope paper was approximately one-fifth of what it would cost to use the larger model for all extractions.

We also took several precautions to ensure that the application updates did not introduce new problems. First, we archived the original factsheets evaluated by the authors to preserve the results reported in this paper. Second, we converted the authors' specific corrections into 17 automated checks. For example, one check verifies that a particular paper is correctly classified as Phase~II. We required each new configuration to pass these checks before deployment. The final configuration passed all 17 checks, whereas each LLM used independently failed one. Third, we tested the complete extraction workflow on samples of 125 and 150 papers, including papers retrieved by the revised searches, before processing the full collection. We also stored the new factsheets separately to compare them with the original results. Finally, we moved the model names, search terms, and field definitions into a single configuration file. The extraction prompts, application filters, and help documentation are generated from this file to ensure consistency when changes are made.

After these updates, the revised searches retrieved 2{,}982 papers, of which 2{,}316 passed the screening step and received a factsheet. These included 244 papers on process monitoring, 1{,}540 on experimental design, and 532 on reliability engineering. The larger model was needed for 37\% of these papers and changed a main classification for 21\%. Both applications now inform users that they have been updated since the initial round of evaluations by the research team and the sample of arXiv authors. Out of respect for the authors' time, we did not ask them to evaluate the revised factsheets again. Hence, we cannot conclude that the introduced changes would result in improved evaluation ratings, despite addressing the issues raised by the arXiv paper authors and passing our automated checks.

\subsection{Practical Limitations and Trade-offs}\label{subsec:limitations}

One issue we have examined only in part is the effect of changing the underlying LLM. For example, we intend to periodically update the model used by \texttt{QE ArXiv Watch} as newer models become available. However, different models may produce different classifications for the same paper, even when the extraction schema remains unchanged \citep{chen2023chatgpt}. Following the arXiv authors' review, we evaluated two replacement models from OpenAI: a smaller model (\texttt{gpt-6-luna}) and a larger model (\texttt{gpt-6.1-sol}). Given identical instructions, the models assigned the same main label to 74\%--98\% of the 125 sampled papers, depending on the field. For applications that accumulate records over time, we suggest checking a sample of previously processed documents when changing models and maintaining a record of which model was used for extraction. Each factsheet in the current version records the models, the schema version, and the extraction date.

The choice of model also depends on the application. In our recall example, the small, locally deployed model performed comparably to the commercial model when extracting three factual fields. We have not evaluated whether the same results would hold for more complex extraction tasks, such as those used in \texttt{QE ArXiv Watch}. Cost is another consideration, particularly when processing large numbers of documents or using VLMs to extract information from scanned PDFs and images. When documents are by themselves very large (eg., reports of 100+ pages), a preprocessing pipeline might be needed to select relevant pages before submitting the document to the LLM or VLM to ensure associated costs are reasonable. Finally, organizations working with proprietary or sensitive information may prefer locally deployed models when sending documents to external providers is not appropriate.

%% file: 07_concluding_remarks.tex
Overall, our objective in this paper was to demonstrate how quality engineers can use LLMs and VLMs to convert unstructured documents into data suitable for further analysis. Through the code examples and two publicly available applications, we show how this can be accomplished without developing specialized NLP pipelines. We believe these tools can reduce the effort required to prepare data for quality engineering applications. Although \texttt{QE ArXiv Watch} monitors research literature, the same architecture of scheduled retrieval, schema-based extraction, and a filterable dashboard can be applied to any recurring stream of documents that must be classified using domain-specific categories, such as supplier corrective action reports, customer complaints, or regulatory updates. However, the responsibility for defining the variables, assessing the quality of the extracted information, and interpreting the resulting data remains with the quality engineer.